\documentstyle[fleqn,12pt]{article}

\newcounter{abc}

\newcounter{Rom}

\newcounter{rom}

\begin{document}

\vspace*{3ex}

\Large
\begin{center}

   \vspace*{2ex}
     {\bf A role of topology and quantum information in physics } \\

   \vspace*{2ex}
                   {\bf Jaroslav HRUBY}

   {\bf Institute of Physics AV CR,Czech Republic }
   {\bf e-mail: jaroslav.hruby@iol.cz  }
\vspace*{3ex}

\begin{abstract}
The role of topology in QIS with physical connection to the
noncommutativity,discretization,supersymmetry, entanglement,
nonseparability and CP violation in physics is discussed.

\end{abstract}

\end{center}

\vspace*{3ex}

\normalsize

\section{Introduction}
\setcounter{equation}{0}

    In recent time it is believed that quantum mechanics (QM) has the
potential to bring about a spectacular revolution in quantum information
science (QIS) [1].

What is more interesting that QIS can give new ideas to QM  and
field theories. There are natural relationships between quantum
entanglements and topological entanglements [2]. The
violation of Bell inequalities by photons more than 10 km is well
established [3]and so from QM the collapse of  physical
state is realized by the superluminal velocity, what is hard to
believe. More interesting explanation is that only quantum
information is nonlocal and topological connected and via
measurement we obtain the collapse to the classical
information.The interesting glue between quantum mathematics and
topology appears as a candidate to show new way how to explain
this old problem.

Topology studies global relationships in spaces, and how one space
can be placed within another, such as knotting and linking of
curves in three-dimensional space.This mathematical area is
popular in physics namely with application of quantum groups. One
way to study topological entanglement and quantum entanglement is
to try making direct correspondences between patterns of
topological linking and entangled quantum states.

A deeper method is to consider braid group representation (BGR) and unitary gates
R that are both universal for quantum computation and are also solutions to the condition for topological
braiding. Such R -matrices are unitary solutions to the Quantum Yang-
Baxter equation (QYBE).

In this way, we can study
the apparently complex relationship among topological entanglement, quantum
entanglement, and quantum computational universality.We can also show how it is
connected with the non-commutative geometry, discretization and supersymmetries.

   The basic question is arising:

    how the field theory,space, topology and information can be connected?

A speculative study was presented by D.Deutsch and myself [4]
and here we discuss another point of view.

It is known non-commutative geometry and quantum groups are of relevance of
space-time quantization and discretization.

   The idea of quantization of space-time using noncommutative
coordinates like
\begin{equation}
  x_{\mu} x_{\nu} -  x_{\nu} x_{\mu} = i\hbar g_{\mu\nu} \ ,
 \ \  x_{\mu} x_{\nu} -  q x_{\nu} x_{\mu} = 0      \label{1.1}
\end{equation}
was presented half century ago .

   For example it is natural to attempt to relate the noncommutativity parameter
\(q<1\) to the minimal uncertainty in length measurement
\begin {equation}
 \delta x>l_{Pl.}= \sqrt{\frac{2\kappa\hbar}{c^3}} \sim 10^{-35}\, m \ ,
     \label{1.2}
\end{equation}
or time measurement
\begin{equation}
 \delta t >\tau_{Pl.}=\frac{l_{Pl.}}{c} \sim 10^{-43}\, s \ .    \label{1.3}
\end{equation}
where $\kappa$, $\hbar$ are the gravitational and Planck constants
and $c$ is the light velocity.

   But this deformation parameter can be obtained
also in the BGR, where b matrices have to satisfy braid relation
[5],[6]:
\begin {equation}
 b_{i}b_{i+1} b_{i}   = b_{i+1}b_{i}b_{i+1} ,1\leq i\leq
 n-1,b_ib_j = b_jb_i, \mid i-j\mid>1,
     \label{1.4}
\end{equation}

while QYBE in R-matrices can be written as follows:

\begin {equation}
 \Re_{i}(x)\Re_{i+1}(xy) R_{i}(y)   = R_{i+1}(y)R_{i}(xy)b_{i+1}(x) ,
     \label{1.5}
\end{equation}

with the asymptotic condition $R(x=0) = b$. The b-matrix and
R-matrix are $n^2 \times n^2$ matrices acting on $V \otimes V$,
where V is an n-dimensional space. As b and R acts on the tensor
product $V_{i} \otimes V_{i+1}$ we denote them $ b_{i}$ and
$R_{i}$, respectively.

 The association of a unitary operator with a braid that respects
 the topological structure of the braid and allows exploration of
 the entanglement properties of the operator.The entanglement
 between two physical states or two-qubit states are known and
 play crucial role in quantum physics and QIS.
 Our aim is to study the geometrical and algebraical fundament of physics and
 QIS. We want to understand the coincidence and try to show the way to the obtaining
 some knowledge about algebraic structure of the physical "space"
 and the connection with quantum information. The algebraic
 structure of the "space" give the possibility of discretization
 and quantization of the space.

 Modern theory of quantum  and braided groups  can be  applied
 in fractional supersymmetries and n-anyonic vector spaces with the generalized
  grassmannian variables ${\theta}^n = 0$, which also can give discretization.

  For example there are possible discretization on the following
bases :
\begin{enumerate}
\item fractional supersymmetry and paragrassmannian q-deformed
      superspace connected with fractional or anyonic statistics
\item on a model, with q-deformed Heisenberg uncertainty relation
      for the null sector[6].
\end{enumerate}

   At this moment is no known basic principle requiring
space or time to be continuous or forbidding limitations on their
units.There is also no known basic principle where is forbidden to
combine physical "space" with information n-qubit Hilbert space
${H_2}^n$. But it is forbidden to transfer physical object via
superluminal velocity like collapse of physical wave function
realize.The known EPR  paradox must be explained more natural way
via collapse of nonlocal quantum information, which can has some
topological fundament.

   The article is organized as follows:

   we start in Sect.\ 2 from
the q-deformed quantum mechanics (QM) and  quantum space time to
obtain quantum space. In  Sect.\ 3 we discuss the quantum
superspace, which can be extended. This is done to show the
possibility for obtaining the richer structure in the fractional
superspace and that the base of the quantization can be done on
the level of such superspace in the general case.
   Superspace is also good example for starting the study an
   algebraic structure of "space". It is well known that the space
   variable for example Bose space variable $x_\mu, \mu=1,...4$ is a
condensate of two Fermi variable
$\overline{\theta}{\gamma}_\mu\theta$ in ordinary
supersymmetries.The algebraic structure of "space" is important
for showing some connections to the QIS.

In Sect.\ 4 we  present basic information about quantum
mathematics and the connection with topological anyonic theory and
QIS is discussed.We also discuss basic information
about supersymmetry and QIS.

In Sect.\ 5 we discuss the QYBE and
universal quantum gate for two-qubit systems.We show the
explanation the entanglement via nonlocal quantum information.
Deformation of entanglement can be applied in physics on every
state/antistate physical system. For example CP discrete symmetry
violation can be explained via noncommutation or equivalently like
$\varphi$ deformation on the braid connecting entangled states.

In Sect.\ 6 we show the application of topological entanglement on
CP violation and in Sect.\ 7 the separability criterion in kaon
system.

\section {Q-deformed quantum mechanics and quantum space-time}

   Limitations on the precision of localization in  spacetime have appeared
in the recent literature as consequence of different approaches
to quantum gravity or q-deformed calculus.
   In studies of quantum group (see for example S.Majid [6]) the commutation
   relation
\begin{equation}
   ab - qba = 0  ,         q\in C
      \label{2.1}
\end{equation}
is among the most typical, together with inhomogeneous relation
\begin{equation}
\tilde{a}\tilde{b}- q'\tilde{b}\tilde{a} = q'' ,
      \label{2.2}
\end{equation}
where ${q',q''}\in C $. For $ q\neq1 $ these Eqs. can be transformed into
another through
\begin{equation}
b = \tilde{b}, a = \tilde{a}+\tilde{b}q''{(q-1)}^{-1} ,  q = q' ,
      \label{2.3}
\end{equation}
Let us remained that the commutation relation of the linear
braided space (see S.Majid [6]) has the form:
\begin{equation}
x_ix_j = x_bx_a{R'^{ab}}_{ij} ,
      \label{2.4}
\end{equation}

   We now show the coincidence between our q-deformed QM and a model of
the discretization of spacetime.

   Let us suppose that $\triangle{x}_{0}\equiv{\tilde{q}}^2-1\approx 0$ is  a  parameter of the
discretization of spacetime and $\tilde{q}$ a parameter of q-deformed QM.

   Let us consider the discretization of standard differential calculus in one
space dimension
\begin{equation}
   [x,dx] = dx\triangle{x}_{0}  ,
      \label{2.5}
\end{equation}
and the action of the discrete translation group
\begin{equation}
   x^ndx = dx(x+\triangle{x}_{0})^n  ,
      \label{2.6}
\end{equation}
\begin{equation}
  \psi(x)dx = dx\psi(x+\triangle{x}_{0}) ,
      \label{2.7}
\end{equation}

for any wave function $\psi$ of the Hilbert space of
QM with the discrete space variable.

   The discrete space variable can be defined as $x=n\triangle x_{0}$, where n is an integer and
$\triangle x_{0}$ is the interval between two discrete space points in this
space variable.

   If we define the derivatives by

\begin{equation}
  d\psi(x) = dx(\partial_{x}\psi)(x)
  (\stackrel{\leftarrow}{\partial_{x}}\psi)(x)dx,
      \label{2.8}
\end{equation}
\begin{equation}
  (\partial_{x}\psi)(x) = \frac{1}{\triangle{x}_{0}}[\psi(x+\triangle{x}_{0})-\psi(x)],
      \label{2.9}
\end{equation}
\begin{equation}
  (\stackrel{\leftarrow}{\partial_{x}}\psi)(x) = \frac{1}{\triangle{x}_{0}}[\psi(x)- \psi(x-\triangle{x}_{0})],
      \label{2.10}
\end{equation}
\begin{equation}
(\stackrel{\leftarrow}{\partial_{x}}\psi)(x) = (\partial_{x}\psi)(x-\triangle{x}_{0})  ,
      \label{2.11}
\end{equation}
then the ordinary one-dimensional  Schr$\ddot{o}$dinger equation will be
\begin{equation}
 \frac{1}{2}\frac{d^2\psi(x)}{dx^2} + [E - U(x)]\psi(x)  = 0,
      \label{2.12}
\end{equation}
with the potential $U(x)$ and wavefunction $\psi(x)\equiv\psi(E,x)$,
corresponding to energy value E,
has on the discrete space the form
\begin{eqnarray}
\lefteqn{ \frac{1}{2({\triangle{x}_{0}})^2}[\psi((n+1)\triangle{x}_{0})-
    2\psi(n\triangle{x}_{0})+ \psi((n-1)\triangle{x}_{0})]+} \nonumber \\
 & + & [E  -  U(n\triangle{x}_{0})]\psi(n\triangle{x}_{0})  = 0 \ .
      \label{2.13}
\end{eqnarray}

We now show the coincidence between such discretization model, noncommutative
differential calculus and q-deformed QM, assuming ${\tilde{q}}^2\approx 1$.

Let us suppose that ordinary continuum space variable y in QM has the
form:
\begin{equation}
y = \lim_{\triangle{x}_{0}\rightarrow0}(1+\triangle{x}_{0})^{\frac{x}{\triangle{x}_{0}}}=
  e^{x}.
\label{2.14}
\end{equation}

   Using Eqs.(38-41) and (44) we get:
\begin{equation}
\partial_{y} = y^{-1}\partial_{x} = (q_{E} + 1)^\frac{-1}{q_{E}}\partial_{x}
\label{2.15}
\end{equation}
   Thus, using $\triangle{x}_{0}\equiv{\tilde{q}}^2-1$, we have
\begin{equation}
  (\partial_{y}\psi)(y) = \frac{\psi((\triangle{x}_{0}+1)y)-\psi(y)}{\triangle{x}_{0}y}=
         \frac{\psi(q^2y)-\psi(y)}{(q^2-1)y}
      \label{2.16}
\end{equation}
\begin{equation}
(\stackrel{\leftarrow}{\partial_{y}}\psi)(y)   =
(\triangle{x}_{0}+1)
\frac{\psi(y)-\psi((\triangle{x}_{0}+1)y))}{\triangle{x}_{0}y} =
\frac{\psi(y)-\psi(q^2y)}{(1-q^{-2})y} \\      \label{2.17}
\end{equation}
what represents derivatives in the differential on the quantum hyperplane .

   We can see that for $\triangle{x}_{0}\rightarrow0$ or ${\tilde{q}}^2\rightarrow1$ we have the ordinary QM and continuous
space time.

 There is unclear in the continuous space-time what is
the ``quantum line'' because classically a coordinate  always
commutes with itself.

   Quite different situation is in the quantum case of the discrete space-time
or of space-time based on grassmannian variables.

\section{Quantum superspace}

   We introduce supersymmetry (SUSY) with superspace \(\{t,\Theta\}\) ,
where $t$ is the time variable and $\Theta$ a Grassmann variable
i.e.\ \(\Theta^{2}=1\).

   We define the supercoordinate
\begin{equation}
 X(t,\Theta)=x_{(0)}(t) +i\,\Theta\, x_{(1)}(t) \ ,          \label{3.1}
\end{equation}
where $x_{(0)}(t)$ is the ordinary  commuting space  coordinate
(Bose or null sector variable) and $x_{(1)}(t)$ is the real
anticommuting variable (Grassmann, Fermi or one-sector).

   The changes of $x_{(0)}(t)$ and $x_{(1)}(t)$ follows from:
\begin{equation}
 \delta X(t,\Theta)=X(t',\Theta')-X(t,\Theta)=i\,\varepsilon\, QX(t,\Theta)
  \ ,                               \label{3.2}
\end{equation}
where SUSY generator
\begin{equation}
 Q=\frac{\partial}{\partial\Theta}+i\,\Theta\frac{\partial}{\partial t} =
   \partial_{\Theta}+i\,\Theta\,\partial_{t}         \label{3.3}
\end{equation}
and $\varepsilon$ is the infinitesimal Grassmann parameter.

   We can see :
\begin{eqnarray}
 \partial X & = & \varepsilon\,\partial_{\Theta}(x_{(0)}
   +i\,\Theta\, x_{(1)}) + i\,\varepsilon\,\Theta\,\partial_{t}(x_{(0)}
   +i\,\Theta\, x_{(1)}) \nonumber \\
& = & i\,\varepsilon\, x_{(1)}+ i\,\varepsilon\,\Theta\,\partial_{t}\,x_{(0)}
  \label{3.4}
\end{eqnarray}
and SUSY transformations for the coordinates $x_{(0)}$ and $x_{(1)}$ :
\begin{equation}
 \delta x_{(0)}=i\,\varepsilon\, x_{(1)} \ ,
\ \ \ \delta x_{(1)}=\varepsilon\,\partial_{t}\,x_{(0)} \ ,
\label{3.5}
\end{equation}

   It follows immediately:
\begin{eqnarray}
 Q^{2} X & = & Q\,[i\,x_{(1)}+i\,\Theta\,\partial_{t}\,x_{(0)}]=
  i\,\partial_{t}\,(x_{(0)}+i\,\Theta\,x_{(1)})=i\,\partial_{t}\,X \ ,
           \label{3.6}  \\
 \mbox{or} & &
 \frac{1}{2} \{Q,Q\} X\equiv H X=i\,\partial_{t}\,X \ , \hspace{15em}
 \label{3.7}
\end{eqnarray}
which suggests that the Hamiltonian of the system be defined as %\\
\(H=\frac{1}{2} \{Q,Q\}\) and the time translation is simply the Hamiltonian
\(H=i\,\partial_{t}\) .

   In this sense the \(N=1\) SUSY (it means one Grassmann $\Theta$)
is the square root of the time translation.

   Let us now to turn to the general case i.e.\ the $F-th$
roots of the time translation \(F=1,2,\ldots\) .

   We need $F$ real Grassmann coordinates \(x_{(j)}(t),\ j=0,1,\ldots,F-1\),
which belong to the following $(F-j)$-sectors $x_{(j)}(t)$ and the
null sector \(x_{(0)}(t)\equiv x(t)\) i.e.\ ordinary coordinate.

   These sectors can be viewed as the components of a quantum superspace
with fractional SUSY .

   We denote fractional quantum superspace

\begin{equation}
X^{F}(t,\Theta) = \sum_{j=0}^{F-1} x_{(j)}(t).{\Theta}^{j}  , \label{3.8}
\end{equation}

where $\Theta$ is a real paragrassmannian variable satisfying
${\Theta}^F=0$ .

   Let us introduce the q-commutation relation

\begin{equation}
 x_{(j)}(t)x_{(F-j)}(t)=q^{j} x_{(F-j)}(t) x_{(j)} \ .       \label{3.9}
\end{equation}

   In this sense the parameter $q^{j} $ connects different sectors.

   Then fractional SUSY has the form:
\begin{eqnarray}
  \delta x_{(j-1)} & = & i\,\varepsilon\,\alpha\,(1-q^{j})\,x_{(j)} \ ,
        \label{3.10a} \\
  \delta x_{(F-1)} & = & \varepsilon\,(F\alpha^{F-1})^{-1}\,\partial_{t}\,
  x_{(0)} \ ,            \label{3.10b}
\end{eqnarray}
 where $\alpha$ is a free constant.

   We have
\begin{equation}
 \delta^{F} x_{(j)}(t)=i^{(F-1)}\varepsilon_{1}\ldots\varepsilon_{F}\
     \partial_{F} x_{(j)}(t)  ,                         \label{3.11}
\end{equation}
since \(\prod\limits_{j=1}^{n}(1-q^{j})=F\)
and
\( \varepsilon\,x_{(j)}(t) = q^{-j}x_{(j)}(t)\,\varepsilon \ \).

   An invariant action is
\begin{equation}
 S= \int dt\, \frac{1}{2}\biggl[(\partial_{t}\,x)^{2}+i\,(F\alpha^{F})\,
   \sum_{j=0}^{F-1}(1-q^{-j})\,(\partial_{t}\,x_{(j)})\,x_{(F-j)}\biggr]
     \label{3.12}
\end{equation}
and fractional SUSY quantum mechanics (SSQM) of order $F$ is defined
through the algebra
 \[Q^{F}=H \ , \ \ [H,Q]=0 \ , \ \ F=2,3,\ldots \ ,\]
where $H$ is the Hamiltonian.

   The fractional SUSY can be extended by the following way:

   For the \(N=2\) SUSY, the superspace is \((t,\Theta_{1},\Theta_{2})\)
and SUSY transformation:
\begin{equation}
 \Theta'_{l}=\Theta_{l}+\varepsilon_{l} \ , \ l=1,2 \ ; \qquad
   t'=t+i\,\varepsilon_{1}\Theta_{1}+i\,\varepsilon_{2}\Theta_{2}
                      \label{3.13}
\end{equation}

Such extended SUSY can has application for anyonic or qubit
superfields \cite{4}.

\section{Some ideas from quantum mathematics}

       A quantum bit (qubit) is a quantum system with a two-dimensional
Hilbert space, capable of existing in a superposition of Boolean states
and of being entangled with the states of other qubits [1].

More precisely a qubit is the amount of the information which is contained
in a pure quantum state from the two-dimensional Hilbert space ${\cal H}_2$.

A general superposition state of the qubit is
\begin{equation}
 |\psi\rangle  = \psi_0 |0\rangle  + \psi_1 |1\rangle  ,
\label{4.1}
\end{equation}
where  $\psi_0$  and   $\psi_1$ are complex numbers, $|0\rangle$
and $|1\rangle$ are kets representing two Boolean states. The
superposition state has the propensity to be a $0$ or a $1$ and
${|\psi_0|}^2 + {|\psi_1|}^2 = 1$.

The eq.(1) can be written as
\begin{equation}
 |\psi\rangle  = {\psi}_0\left( \begin{array}{c} 1 \\ 0 \end{array} \right)
 + {\psi}_1 \left( \begin{array}{c} 0 \\ 1 \end{array} \right)  ,
\label{4.2}
\end{equation}

where we labelled $\left( \begin{array}{c} 1 \\ 0 \end{array}
\right) $ and $\left( \begin{array}{c} 0 \\ 1 \end{array} \right)$
two basis states zero and one.

The Clifford algebra relations of the $2\times2$ Dirac matrices is
\begin{equation}
\left\{{\gamma}^\mu,{\gamma}^\nu \right\} = 2 {\eta}^{\mu\nu} ,
\label{4.3}
\end{equation}
where
\begin{equation}
 {\eta}^{\mu\nu}    =  \left( \begin{array}{cc}
                              -1 & 0 \\
                               0 & 1 \end{array} \right).
\label{4.4}
\end{equation}

We choose the representation
\begin{equation}
 {\gamma}^0 =i{\sigma}^2   =  \left( \begin{array}{cc}
                              0 & 1 \\
                             -1 & 0 \end{array} \right)
\label{4.5}
\end{equation}
and
\begin{equation}
 {\gamma}^1 ={\sigma}^3   =  \left( \begin{array}{cc}
                              1 & 0 \\
                              0 &-1  \end{array} \right) ,
\label{4.6}
\end{equation}
where $\sigma$ are Pauli matrices and ${\gamma}^5={\gamma}^0{\gamma}^1$ .

 The projectors have the form
\begin{equation}
P_0 = \frac{1+{\gamma}^1}{2} =  \left( \begin{array}{cc}
                              1 & 0 \\
                              0 & 0  \end{array} \right)
 \label{4.7}
\end{equation}
and
\begin{equation}
P_1 = \frac{1-{\gamma}^1}{2} =  \left( \begin{array}{cc}
                              0 & 0 \\
                              0 & 1  \end{array} \right)
 \label{4.8}
\end{equation}

These projectors project qubit on the basis states zero
and one:
\begin{equation}
P_0  |\psi\rangle  = {\psi}_0\left( \begin{array}{c} 1 \\ 0
\end{array} \right), P_1  |\psi\rangle  = {\psi}_1\left(
\begin{array}{c} 0 \\ 1 \end{array} \right) \label{4.9}
\end{equation}
and represent the physical measurements - the transformations of qubits to the
classical bits.

In classical information theory the Shannon entropy is well defined :
\begin{equation}
S_{CL}(\Phi) =-\sum_{\phi} p(\phi) {\log}_2 p(\phi), \label{4.10}
\end{equation}
where the variable $\Phi$ takes on value $\phi$ with probability  $p(\phi)$
and it is interpreted as the uncertainty about $\Phi$.

The quantum analog is the von Neumann entropy $S_{q}({\rho}_{\Psi})$ of a quantum
state $\Psi$ described by the density operator ${\rho}_{\Psi}$:
\begin{equation}
S_Q(\Psi) = - {Tr}_{\Psi}\left[ {\rho}_{\Psi}
{\log}_2{\rho}_{\Psi} \right], \label{4.11}
\end{equation}
where ${Tr}_{\Psi}$  denotes the trace over the degrees of freedom associated
with $\Psi$.
x
The von Neumann entropy has the information meaning, characterizing (asymptotically)
the minimum amount of quantum resources required to code an ensemble of quantum
states.

The density operator ${\rho}_{\psi}$ for the qubit state $ |\psi\rangle $ in (1)
is given:

\begin{equation}
\rho = |\psi\rangle \langle\psi| = {|\psi_0|}^2|0\rangle \langle0|
+ \psi_0{\psi_1}^* |0\rangle \langle1| + {\psi_0}^*\psi_1
|1\rangle \langle0| +{|\psi_1|}^2|1\rangle \langle1| \label{4.12}
\end{equation}
and corresponding density matrix is
\begin{equation}
{\rho}_{kl} = \left( \begin{array}{cc}
                              {|\psi_0|}^2 & \psi_0{\psi_1}^*  \\
                           {\psi_0}^*\psi_1 & {|\psi_1|}^2  \end{array} \right)
\label{4.13}
\end{equation}
and $k,l=0,1.$

The von Neumann entropy reduces to a Shannon entropy if ${\rho}_{\Psi}$ is a mixed
state composed of orthogonal quantum states.

A quantum gate is a the analog of a logic gate in a classical computer.The NOT gate is
$X|k\rangle = |k \oplus 1\rangle $ , where the addition is mod(2).
The unitary quantum gate defined

\begin{equation}
                   M_{-}  =  \left( \begin{array}{cc}
                              0 & 1 \\
                              1 & 0  \end{array} \right)
 \label{4.14}
\end{equation}

defines an action $ M_- |0\rangle \ = |1\rangle \ $, $ M_- |1\rangle \ = |0\rangle \ $
is called a quantum not-gate.

The matrix
\begin{equation}
            \sqrt {M_{-}}  =  \left( \begin{array}{cc}
                              \frac{1+i}{2} & \frac{1-i}{2} \\
                              \frac{1-i}{2} & \frac{1+i}{2}   \end{array} \right)
 \label{4.15}
\end{equation}

If we denote $|0\rangle\ =  {(1,0)}^T$ and   $|1\rangle \ =  {(0,1)}^T$
the action of

\begin{equation}
\sqrt {M_{-}}|0\rangle  =  \frac{1+i}{2}|0\rangle +
\frac{1-i}{2}|1\rangle
 \label{4.16}
\end{equation}
and
\begin{equation}
\sqrt {M_{-}}|1\rangle  =  \frac{1-i}{2}|0\rangle + \frac{1+i}{2}
|1\rangle
 \label{4.17}
\end{equation}

In general, an $N$-qubit can be in an arbitrary superposition of
all $2^{N}$ classical states:
\[ |\psi_{N}\rangle=\sum\alpha_{x}|x\rangle \ , \
     x\in\{|0x\rangle,|1\rangle\}^{N}\
           \mbox{and}\ \sum_{x}|\alpha_{x}|^{2}=1 \ . \]

It is known that two bit gates are universal
for quantum computation, which is likely to greatly simplify the
technology required to build quantum computers.

Unitary gates, which play crucial role in QIS, are connected with
R-matrices from QYBE and via Yang-Baxterization Hamiltonians can
be expressed as square root of b-matrices. This established the
connection between topology and QM, especially it gives
topological interpretation of entanglement. Here we show it on the
two-qubit state.

To understand this we start with a simple idea of quantum
algorithm square root of not
 via partner--superpartner case.
Supersymmetry is special case of anyonic Lie algebra
$C[\theta]/{\theta}^n$ with one coordinate $\theta ,
{\theta}^n=0$(for our Grassmann variable ${\Theta}^2 = 0$)\cite{6}
with anyonic variables fulfil:
\begin{equation}
\theta''=\theta +\theta'  ,  \theta'\theta = e^{\frac{2\pi
i}{n}}\theta\theta'  \label{4.18}
\end{equation}
 and the category of n-anyonic vector spaces where objects are
 $Z_n$-graded spaces with the braided transposition:

\begin{equation}
\Psi(x \otimes y)= e^{{\frac{2\pi i}{n}\mid x\parallel y\mid}}y
\otimes x  \label{4.19}
\end{equation}
 on elements $x,y$ of homogenous degree $\parallel$.

 Thus an anyonic braided group means as $Z_/n$-graded algebra B
 and coalgebra defined as $\varepsilon$:
 \begin{equation}
 B\rightarrow C , \triangle(x)=x_A\otimes x^A ,\label{4.20}
\end{equation}
where summation understood over terms labelled by  A of
homogeneous degree.

Coassociative and counital is in the sense:
 \begin{equation}
 x_{AB}\otimes {x_A}^B \otimes x^A = x^A\otimes {X^A}_B\otimes x^{AB} ,
 \varepsilon(x_A)x^A =x = x_A\varepsilon (x^A) \label{4.21}
\end{equation}
and obeying
 \begin{equation}
 (xy)_A\otimes (xy)^A = x_Ax_B\otimes x^Ax^Be^{{\frac{2\pi i}{n}\mid x^A\parallel
 x_B\mid}}, \label{4.22}
\end{equation}
for all ${x,y}\in B$.The axioms for the antipode are as for usual
quantum groups \cite{6}. There is shown that anyonic calculus and
anyonic matrices are generalization of quantum matrices and
supermatrices.

The $N^2$ generators ${t^i}_j= f(i)-f(j)$, where f is a degree
$Z_/n$ associated with the row or column fulfil
\begin{equation}
e^{(\frac{2\pi i}{n})\{f(i)f(k)+f(j)f(l)\}}{{{\Re^i}_a}^k}_b
{t^a}_b {t^b}_l=e^{(\frac{2\pi i}{n})\{f(j)f(l)+f(i)f(k)\}}{t^k}_b
{t^i}_a{{{\Re^a}_j}^b}_l, \label{4.23}
\end{equation}

\begin{equation}
\triangle {t^i}_j = {t^i}_a\otimes {t^a}_j, \label{4.24}
\end{equation}

\begin{equation}
\varepsilon {t^i}_j = {{\delta}^i}_j \label{4.25}
\end{equation}

 It is required that $\Re$ obeys certain anyonic QYBE.
 One method to obtain $\Re$ is start with certain unitary solution
 R of the usual QYBE and "transmute: them.
 In the diagrammatic notation this braided mathematics is known \cite
 {5}. In QIS a "wiring" notation (which is known in physics like Feynman
 diagrams) was used for information flows for example for
 teleportation \cite{4}.
 In the diagrammatic notation we "wire" the outputs of maps into
 the inputs of other maps to construct the algebraic operation
Information flows along these wires except that under and over
crossing are nontrivial operators , say $U$ and $U^{-1}$. Such
operators can be universal quantum gates in QIS. Generally in
anyonic we have new richer kind of "braided quantum field
information mathematics".

To show it we begin  with coincidence of the supersymmetric square
root in SSQM $n=2$ and square root of the not gate in QIS as an
illustrative example.

It is well known SSQM is generated by supercharge operators
$Q^{+}$ and \(Q^{-}=(Q^{+})^{+}$ which together with the
Hamiltonian $H=2H_{SSQM}$ of the system, where

\begin{eqnarray*}
 H_{SSQM} & = & \frac{1}{L}\left(
     \begin{array}{cc}
       -\frac{d^{2}}{d\,t^{2}}+v^{2}+v' & 0 \\
       0 & -\frac{d^{2}}{d\,t^{2}}+v^{2}-v'
     \end{array}  \right)    \\
 H & = & \left(
     \begin{array}{cc}
       H_{0} & 0 \\ 0 & H_{1}
     \end{array}  \right) =
   \left(
     \begin{array}{cc}
       A^{+}A^{-} & 0 \\ 0 & A^{-}A^{+}
     \end{array}  \right) =
     -\left( \frac{d^{2}}{dx^{2}}\right)+\sigma_{3}v'
\end{eqnarray*}

fulfil the superalgebra

\begin{equation}
{Q^{\pm}}^{2}=0 \ , \ \ [H,Q^{-}]=[H,Q^{+}] \ ,
                              \ \ H=\{Q^{+},Q^{-}\}=Q^{2}\, \label{4.26}
\end{equation}

where

\[ Q^{-}=\left(
     \begin{array}{cc}
       0 & 0 \\  A^{-} & 0
     \end{array}  \right) \ , \
 Q^{+}=\left(
     \begin{array}{cc}
       0 & A^{+} \\  0 & 0
     \end{array}  \right)   \ , \  \  Q = Q^{+}+Q^{-} \]
and

\begin{equation}
 A^{\pm}=\pm\frac{d}{dx}+v(x) \ , \ \ v'=\frac{dv}{dx}  . \label{4.27}
\end{equation}

Such Hamiltonians \(H_{0},H_{1}\) fulfil
\begin{equation}
H_{0}A^{+}=A^{+}H_{1} \ , \ \ A^{-}H_{1}=H_{0}A^{-}.\label{4.28}
\end{equation}
The eigenfunctions of the Hamiltonian
\begin{equation}
H  =  \left(     \begin{array}{cc}
       H_{0} & 0 \\ 0 & H_{1}
     \end{array}  \right)
\end{equation}

 are  $\phi=(|0\rangle,|1\rangle)^T$ and then

\begin{equation}
     Q^0\phi=|1\rangle,
\end{equation}

\begin{equation}
     Q^1\phi=|0\rangle.
\end{equation}

 where we denote $Q^-=Q^0$and$Q^+=Q^1$ ,respectively.

The previous relations in lead to the double degeneracy of all positive
energy levels, of belonging to the ``0'' or ``1'' sectors specified
by the grading state operator $S=\sigma_{3}$, where
\[[S,H]=0 \ \mbox{and}\ \{S,Q\}=0 \ . \]

The $Q$ operator transforms eigenstates with \(S=+1\), i. e. the null-state
$|0\rangle$  into eigenstates with \(S=-1\), i. e. the one-state
$|1\rangle$ and vice versa.

With this notation the square root of not gate $M_- = \sqrt M_-
\sqrt M_-$ is represented by the unitary matrix $\sqrt M_-$:
\[\sqrt M_- =\frac{1}{2}\left[
     \begin{array}{cc}
       1+i & 1-i \\  1-i & 1+i
     \end{array}  \right] \ , \]
that solves:
\begin{equation}
  \sqrt M_- \sqrt M_- |0\rangle =\sqrt M_- (\frac{1+i}{2}|0\rangle + \frac{1-i}{2} |0\rangle)  = |1\rangle ,
    \end{equation}
\begin{equation}
  \sqrt M_- \sqrt M_- |1\rangle  =  |0\rangle
\end{equation}

In such way this supersymmetric double degeneracy represents two level
quantum system and we can see the following:

\begin{eqnarray*}
 Q & = & Q^{+}+Q^{-}= \left(
     \begin{array}{cc}
       0 & A^{+} \\  A^{-} & 0
     \end{array}  \right) \ , \\
  \tau & = & \sigma_{3} = \left(
     \begin{array}{cc}
       1 & 0  \\  0 & -1
     \end{array}  \right) \ , \\
 \{Q,\tau\} & = & 0 \ .
\end{eqnarray*}

It implies that the operator supercharge $Q$ really transforms the
state \(|0\rangle,\ |1\rangle\) as the operator square root of not
quantum algorithm operator $M_-$. In such a way supersymmetric
``square root'' corresponds the ``square root of not'' in QIS. We
can ask if generally the Hamiltonians of the unitary braiding
operators which leads to the Schr\"{o}dinger equations are square
root of QIS unitary gates. Such QIS unitary gates are unitary
solutions of QYBE. The answer is positive and it can be explicitly
shown for two-qubit system.

   A system of two quantum bits is four dimensional space $H_4 = H_2 \otimes H_2$
having orthonormal basis $|00\rangle, |01\rangle, |10\rangle,
|11\rangle $ .

    A state of two-qubit system is a unit-length vector
\begin{equation}
  {a}_0 |00\rangle  + {a}_1 |10\rangle + {a}_2 |01\rangle  + {a}_3 |11\rangle  ,
\label{1.1}
\end{equation}
so it is required
${|{a}_0|}^2 + {|{a}_1|}^2 + {|{a}_2|}^2 + {|{a}_3|}^2  = 1$.

We can see a state $z\in H_4$
\begin{equation}
  z=\frac{1}{2}( |00\rangle  + |01\rangle +  |10\rangle  + |11\rangle )= \frac{1}{\sqrt{2}}(|0\rangle + |1\rangle)\frac{1}{\sqrt{2}}(|0\rangle + |1\rangle) ,
\label{1.1}
\end{equation}
of a two-qubit system is decomposable, because it can be written
as a product of states in $H_2$. A state that is not decomposable
is entangled.

 Consider the unitary matrix
\begin{equation}
 \overline{R}= \left( \begin{array}{cccc}
                              a_0 & 0 & 0 & 0\\
                              0 & 0 & a_3 & 0\\
                              0 & a_2 & 0 & 0\\
                              0 & 0   & 0 & a_1   \end{array} \right) ,
\label{1.1}
\end{equation}

defines a unitary mapping, whose action on the two-qubit basis is
\begin{equation}
    \Psi=\overline{R}(\psi\otimes\psi) =       \overline{R} \left( \begin{array}{c}
                              |00\rangle\\
                              |01\rangle\\
                              |10\rangle\\
                              |11\rangle  \end{array}\right) = \left( \begin{array}{c}
                              a_0|00\rangle\\
                              a_3|10\rangle\\
                              a_2|01\rangle\\
                              a_1|11\rangle  \end{array}\right) . \label{1.2}
\end{equation}

For $a_0a_1 \neq a_2a_3$ the state is entangled.For example the state is
entangled $\frac{1}{\sqrt{2}}(|10\rangle +|01\rangle)$ is
entangled.

\section{The QYBE and universal quantum gate }

    Matrix
\begin{equation}
 M_{CNOT}=                 \left( \begin{array}{cccc}
                              1 & 0 & 0 & 0\\
                              0 & 1 & 0 & 0\\
                              0 & 0 & 0 & 1\\
                              0 & 0 & 1 & 0  \end{array} \right) ,
\label{1.1}
\end{equation}
defines a unitary mapping, whose action on the two-qubit basis is

\begin{equation}
 M_{CNOT} \left( \begin{array}{c}
                              |00\rangle\\
                              |01\rangle\\
                              |10\rangle\\
                              |11\rangle  \end{array}\right) = \left( \begin{array}{c}
                              |00\rangle\\
                              |01\rangle\\
                              |11\rangle\\
                              |10\rangle  \end{array}\right) . \label{1.2}
\end{equation}
    Gate $M_{CNOT}$ is called controlled not, since the target qubit
    is flipped if and only if the control bit is 1.

 Let R
\begin{equation}
 R   = \frac{1}{\sqrt{2}} \left( \begin{array}{cccc}
                              1 & 0 & 0 & 1\\
                              0 & 1 &-1 & 0\\
                              0 & 1 & 1 & 0\\
                              1 & 0 & 0 & 1  \end{array} \right) ,
\label{1.6}
\end{equation}
be the unitary solution to the QYBE.

Let $M = M_1\bigotimes M_2$, where
\begin{equation}
 M_1   =  \frac{1}{\sqrt{2}}\left( \begin{array}{cc}
                              1 & 1\\
                              1 &-1  \end{array} \right) ,
\label{1.1}
\end{equation}
and
\begin{equation}
 M_2   =  \frac{1}{\sqrt{2}}\left( \begin{array}{cc}
                              -1 & 1\\
                               i & i  \end{array} \right) ,
\label{1.7}
\end{equation}

Let $N = N_1\bigotimes N_2$, where
\begin{equation}
 N_1   =  \frac{1}{\sqrt{2}}\left( \begin{array}{cc}
                              1 & i\\
                              1 &-i  \end{array} \right) ,
\label{1.1}
\end{equation}
and
\begin{equation}
 N_2   = - \frac{1}{\sqrt{2}}\left( \begin{array}{cc}
                              1 & 0\\
                              0 & i  \end{array} \right) ,
\label{1.7}
\end{equation}

Then $M_{CNOT} = M\cdot R\cdot N$ can be expressed in terms of R.
 In the QYBE solution R-matrices usually depends on the
 deformation parameter q and the spectral parameter x. Taking the
 limit of $x\longrightarrow0$ leads to the braided relation from
 the QYBE and the BGR b-matrix from the R-matrix. Yang-
 Baxterization is construct the $R(x)$ matrix from a given BGR
 b-matrix.
    The BGR b of the eight-vertex model assumes the form
\begin{equation}
 b\pm   = \left( \begin{array}{cccc}
                              1 & 0 & 0 & q\\
                              0 & 1 & \pm1 & 0\\
                              0 & \mp1 & 1 & 1\\
                        -q^{-1} & 0    & 0 & 1  \end{array} \right) , \label{1.1}
\end{equation}

    It has two eigenvalues $\Lambda_{1,2}= 1\pm i $. The
    corresponding $R(x)$-matrix via Yang-Baxterization is obtained
    to be
    \begin{equation}
 R_{\pm}(x) = b + x \Lambda_1 \Lambda_2 =
 \left( \begin{array}{cccc}
                              1+x & 0 & 0 & q(1-x)\\
                              0 & 1+x & \pm(1-x) & 0\\
                              0 & \mp(1-x) & 1+x & 1\\
                    -q^{-1}(1-x)& 0    & 0 & 1+x  \end{array} \right),
\label{1.1}
\end{equation}

    Introducing the new variables $\theta,\varphi$ as follows
\begin{equation}
\cos\theta = \frac{1}{\sqrt{1+x^2}}, \sin \theta =
\frac{x}{\sqrt{1+x^2}}, q = e^{i\varphi}\label{1.1}
\end{equation}

the R-matrix has the form

\begin{equation}
R_\pm (\theta) = \theta\cos(\theta)b_{\pm}(\varphi) +
\sin(\theta)b_{\pm}^{-1}(\varphi)\label{1.1}
\end{equation}

The time-independent Hamiltonian $H_\pm$ has the form
\begin{equation}
 H_\pm = -\frac{i}{2}{b_\pm}^2(\varphi) = \frac{i}{2} \left( \begin{array}{cccc}
                              0 & 0 & 0 & -e^{i\varphi}\\
                              0 & 0 & \mp1 & 0\\
                              0 & \pm1 & 0 & 0\\
                        e^{-i\phi} & 0    & 0 & 0  \end{array} \right) , \label{1.1}
\end{equation}

\section{Topology and entanglement in physical application}

The braid group representation $b_\pm(\varphi)$ yield the Bell
states with the phase factor $e^{i\varphi}$

\begin{equation}
 b_\pm(\varphi) \left( \begin{array}{c}
                              |00\rangle\\
                              |01\rangle\\
                              |10\rangle\\
                              |11\rangle  \end{array}\right) =\frac{1}{\sqrt{2}} \left( \begin{array}{c}
                              |00\rangle + e^{i\varphi}|11\rangle\\
                              |10\rangle \pm |01\rangle\\
                              \mp|01\rangle + |10\rangle\\
                              -e^{-i\varphi}|00\rangle + |11\rangle \end{array}\right) . \label{1.2}
\end{equation}

which shows that $\varphi=0$ leads to the Bell states, the maximum
of entangled states:

\begin{equation}
\frac{1}{\sqrt{2}}(|00\rangle \pm |11\rangle),
\end{equation}
\begin{equation}
\frac{1}{\sqrt{2}}(|10\rangle \pm |01\rangle).
\end{equation}

In this chapter we want to show  possible explanation of
experimental measurements of the CP violation, using  results from
the QIS presented here.

For this purpose we consider in the  $K\bar{K}$  systems entangled
 states in $H_4$:
\begin{equation}
   |\Phi_1\rangle = \frac{1}{\sqrt{2}}( \mid KK\rangle  + \mid \bar K \bar K\rangle ) , \label{1}  \\
   |\Phi_2\rangle = \frac{1}{\sqrt{2}}( \mid K K\rangle  - \mid \bar K \bar K\rangle ) , \label{2}  \\
\end{equation}
and

\begin{equation}
   |\Phi_3\rangle = \frac{1}{\sqrt{2}}( \mid K\bar{K}\rangle  + \mid \bar K K\rangle ) , \\
   |\Phi_4\rangle = \frac{1}{\sqrt{2}}( \mid K\bar{K}\rangle  - \mid \bar K K\rangle ) , \label{4}  \\
\end{equation}

The necessary codefinition of strangeness $\hat{S}$ and CP can be
rigorously justified by applying local realism to a kaon belonging
to a correlated kaon pair.

From this point of view the experimental measurement of
$K\rightarrow 2\pi$ through the quantity
\begin{equation}
\eta_f  \equiv  \frac{A(\mbox{entagled L and S states}\rightarrow
f)}{A(S\rightarrow f)},
\end{equation}

where f is a pair of either two charged or two neutral pions,
can be the effect from
the measurement of the quantum nonseparability of the entangled
kaon states and the discrete symmetry CP is preserved.

The above classical information analysis $K\bar{K}$ system via
probabilities as is usual is not sufficient for the description of
the quantum nonseparability.

The quantum nonseparability in the $K\bar{K}$ system needs to
study this quantum system via the conditional density matrix and
correlations between kaon states in QIS.

In the case of $\Phi$  and B-factories, where the neutral meson
states produced ( $K\bar{K}$ and  $B\bar{B}$, respectively)
constitute correlated Einstein-Podolsky-Rosen states (EPR) , the
knowledge that one of the two mesons decays at a given time
through a charge specific channel ("tagging $\pm$ ") allows one to
unambiguously infer the charge of the accompanying meson state at
the same time.

If the separability criterion in kaon system is in the sense that
in every moment we have correlation in every kaon system as two-
qubit system in every moment ( it means that every $|K\rangle$ is
correlated with $|K\rangle$ or $|\bar K\rangle$), then we have
four states $\Phi_ {1,2,3,4}$ with "tagging" charges CP and S
$\pm$ .

It gives the possibility to explain CP-violation via deformation
of entanglement (it means Bell-states $\Phi _{1,2,3,4} $ are not
maximally entangled)

The violation can be realized via the phase factor $e^{i\varphi}$
\begin{equation}
 b_\pm(\varphi) \left( \begin{array}{c}
                              |KK\rangle\\
                              |K\bar K\rangle\\
                             |\bar KK\rangle\\
                              |\bar K\bar K\rangle  \end{array}\right) =\frac{1}{\sqrt{2}} \left( \begin{array}{c}
                              |KK\rangle + e^{i\varphi}|\bar K\bar K\rangle\\
                              |\bar KK\rangle \pm |K\bar K\rangle\\
                              \mp|K\bar K1\rangle + |\bar KK\rangle\\
                              -e^{-i\varphi}|KK\rangle + |\bar K\bar K\rangle \end{array}\right) . \label{1.2}
\end{equation}

 We can see two possibilities:

 one of the  possible explanation is a topological origin of CP violation and the
 second possibility is the separability criteria in kaon system.

\section{Separability criterion in kaon system}

We consider a pair of $CP =\pm 1$ , $ \mid S\rangle$ and $\mid
L\rangle$ in an impure mixing of pure entangled states ,
consisting a fraction $\lambda$.

We shall introduce the following pure  $CP = \pm1$ entangled states:
\begin{equation}
\mid SL\rangle = \frac{1}{\sqrt{1+
\mid\varepsilon\mid^2}}\mid S\rangle -
\frac{\bar{\varepsilon}}{\sqrt{1+ \mid\varepsilon\mid^2}}\mid
L\rangle
\end{equation}
and pure $\hat{S} = \pm1 $  entangled states:
\begin{equation}
\mid S\bar{S}\rangle , \mid
L\bar{L}\rangle ,
\end{equation}
where $\varepsilon$ is a little complex parameter, which define a
influence of the second state, which was initially present. Let us
now consider a mixed states with a fraction parameter $\lambda$,
i.e. a real number between 0 and 1:
\begin{equation}
\rho(\varepsilon,\lambda)= \lambda P _{\mid SL\rangle} +
\frac{1}{2}(1-\lambda) (P_{\mid SS\rangle} + P_{\mid LL\rangle} ),
\end{equation}
Using the Horodecki notations \ cit {8}:
\begin{equation}
\rho(\varepsilon,\lambda) = \frac{1}{4}[ {\bf I} +
\lambda\frac{1-\mid\varepsilon\mid^2}{1+\mid\varepsilon\mid^2}(\sigma_{z}\otimes{\bf
I} - {\bf I}\otimes\sigma_{z})
+(1-2\lambda)\sigma_{z}\otimes\sigma_{z} -
2\lambda\frac{\mid\varepsilon\mid}{1+\mid\varepsilon\mid^2}
(\sigma_{x}\otimes\sigma_{x} + \sigma_{y}\otimes\sigma_{y}),
\end{equation}
where $\sigma$ are Pauli matrices. Following \ cite {8} $
\rho(\varepsilon,\lambda) $ can violate Bell inequality if
\begin{equation}
M(\rho(\lambda,\varepsilon)) = \mbox{max} \{(2\lambda-1)^2 +
4{\lambda}^2\left(\frac{\mid\varepsilon\mid}{1+\mid\varepsilon\mid^2}\right)^2
, 8
\lambda^2\left(\frac{\mid\varepsilon\mid}{1+\mid\varepsilon\mid^2}\right)^2
\} > 1.
\end{equation}
So for
\begin{equation}
\lambda > \frac{1}{2}(1 -
\frac{\mid\varepsilon\mid}{1+\mid\varepsilon\mid^2})^{-1},
\end{equation}
the quantum density matrix has the negative determinant, and
therefore a negative eigenvalue.It is the inseparability
criterion.

We can see that for $\varepsilon = 0$ there is no CP violation and
in the  experiments  is measured rather the quantum
nonseparability in the $K\bar{K}$ system.

It is easily seen that the fraction  $\frac{1}{2}(1-\lambda) $
connected with the $\mid S(t)\bar{S}(0)\rangle$ can have the
experimental value for $ \eta_{2\pi}\sim 2.27\times10^{-3}$. From
this the value for the fraction $\lambda$ follows :
\begin{equation}
\lambda  =  0.99546  > \frac{1}{2},
\end{equation}
 and  this experimental value for $\lambda$ is in agreement with
the nonseparability of the kaon states.

Of course we can use more complicated models simulating the
contamination of the purely fully entangled states by random
sources.We use two parameters: $\alpha$ allowing admixture of
single L and S kaons and $v$ for pairs of entangled kaon states.

 The contamined source is given by
\begin{equation}
\rho(LS) = \alpha[v \rho_E +(1-v)\rho_{R}] +
(1-\alpha)[\frac{\rho_{L,S}(t)\rho_V-\rho_V\rho_{L,S}(0)}{2}],
\end{equation}
where the idealized source with maximally entangled state E is
\begin{equation}
\rho_E = \mid E\rangle \langle E \mid, \mid E\rangle = \frac{1}{2}
(\mid S(t)\bar{S}(0)\rangle -  \mid L(t)\bar{L}(0)\rangle )
\end{equation}
the random source of L and S states
\begin{equation}
\rho_R = \frac{1}{4}( \mid S\bar{S}\rangle \langle \bar{S}S\mid +
\mid SL\rangle \langle LS\mid  + \mid LS\rangle \langle SL\mid +
\mid L\bar{L}\rangle \langle \bar{L}L\mid ).
\end{equation}
The random source for single L and S states is
\begin{equation}
\rho_{L,S} = (\mid S\rangle \langle\bar{S}\mid + \mid L\rangle
\langle\bar{L}\mid ),
\end{equation}
and the vacuum term is
\begin{equation}
\rho_V = \mid 0\rangle \langle 0\mid,
\end{equation}
the same for the proper time t and time 0.

Direct evaluation of the matrix element of the matrix elements of
$\rho$  and then diagonalizing it gives for the entropy of the
composite kaon system :
\begin{equation}
- S_{Q}(LS) = \frac{3}{4}\alpha(1-v)\ln{\frac{\alpha(1-v)}{4}}    \\
+  \frac{1}{4}\alpha(1+3v)\ln{\frac{\alpha(1+3v)}{4}} +
\alpha(1-\alpha)\ln{\frac{(1-\alpha)}{4}}
\end{equation}
and the entropy for the single kaon system
\begin{equation}
- S_{Q}(L,S) =  \frac{1+\alpha}{2}\ln{\frac{(1+\alpha)}{4}} +
\frac{1-\alpha}{2}\ln{\frac{(1-\alpha)}{4}}
\end{equation}
is the same for the proper time t and time 0.

The boundary criterion for the entanglement is obtained by
equating both entropies $S_Q(LS)=S_Q(L,S)$ and inseparability
criterion is given [4]:
\begin{equation}
 \alpha v > \frac{1}{\sqrt2}.
\end{equation}

If we have no single states $\alpha = 1$ the parameter $v=\lambda$
 gives us the information about the nonseparability of kaon
entanglement pairs. For the parameter $\alpha$ we get the value
\begin{equation}
 \alpha > 071033
\end{equation}
and for this value there is  the nonseparability of the single
kaon states.

\section{Conclusions}

The new development in QIS and appearing the topological fundament
of entanglement.
 The concept of entanglement for pure quantum states was
established in the early days of quantum mechanics, but the
crucial role in QIS gives quite new interpretation. We hope that
the geometrical and topological point of view on QIS can give
quite new theories for physics.

Quite new is topological understanding of the violation of famous
effect of the violation the CP discrete symmetry in physics.

\end{document}